\documentclass[12pt]{article}
\usepackage[dvips]{graphicx}
\usepackage{wrapft}
\def\be{\begin{eqnarray}}
\def\ee{\end{eqnarray}}

\def\abstract#1{\vskip 7mm 
\begin{center}{\large Abstract}\par \bigskip
\begin{minipage}[c]{12cm}
\small #1
\end{minipage}
\end{center}
}
\def\title#1{\begin{center}{\Large\bf #1}\end{center}}
\def\author#1{\vskip 5mm \begin{center}{#1}\end{center}}
\def\address#1{\begin{center}{\it #1}\end{center}}
\newcommand{\bfr}{\begin{flushright}}
\newcommand{\efr}{\end{flushright}}
\begin{document}
\vspace*{-0cm}
\title{An Infinite Number of Static Soliton Solutions to the 5D Einstein-Maxwell Equations with a Dilaton Field}
\author{Takahiro AZUMA\footnote{E-mail: azuma@dokkyo.ac.jp} 
and Takao KOIKAWA\footnote{E-mail: koikawa@otsuma.ac.jp}
}
\vspace{1cm}
\address{
${}^1$ Faculty of International Liberal Arts, Dokkyo University, \\ 
Soka 340-0042, Japan \\
${}^2$ School of Social Information Studies, Otsuma Women's University, \\
Tama 206-8540, Japan
}
\vspace{3.5cm}
\abstract{
We study the 5D static Einstein-Maxwell equations with a dilaton field. We  construct an infinte number of solutions by using a soliton technique. We study the rod structure of 2-soliton solution and show the 5D dilatonic black ring and black hole solutions are included. 
}

\newpage
\setcounter{page}{2}
\noindent
%%%%%%%%%%%%%%%%%%%%%%%%%%%%%%%%%%%%%%%%%
%
%    section 1
%
%%%%%%%%%%%%%%%%%%%%%%%%%%%%%%%%%%%%%%%%%
\section{Introduction}

Ever since the discorvery of the black ring solution by Emparan and Reall\cite{ER}, much efforts have been made to find the solutions in five dimensions. The 5D black holes exhibit qualitative new properties in contrast with the siblings in four dimensions. Emparan and Reall made use of the C-metric, that enabled them to obtain their paticular solution. After that, many solution generating techniques have been developed\cite{Pom}-\cite{EF}. One of them is to use the inverse scattering method(ISM)\cite{BS}, and the symmetries of the theory after the dimensional reduction are also made use of in genelating new solutions. One of the authors applied ISM to the 5D static Einstein equation\cite{K}, and then we also applied this method to the 5D stationary Einstein equation\cite{AK1}. The 5D rotating black ring solution is the solution in  which the gravity and cetrifigual forces cancel. We expected to construct the solution to the 5D Einstein-Maxwell(EM) equations in which the gravity and electric repulsive forces cancel in the static case, and found a charged black ring solution with a conical singularity\cite{AK2}. In these series of papers, we have shown an infinite number of solutions which shows that these equations are completely integrable systems, as in the 4D case\cite{AK4D1}-\cite{AK4D5}. Our conjecture is that the black hole solutions are given by the soliton solutions also in the 5D case. 

The study of 5D stationary Einstein equation with electric and dilatonic charges, which is inspired from the string theory and the brane-world scenario with large extra dimensions, is intereresting, but the solution has not been found yet. As an first step forward it, we study the 5D static case.
The purpose of the present paper is to study the 5D Einstein-Maxwell-dilaton(EMd) equations and show its complete integrability by explicit construction of an infinite number of solutions. We study the 2-soliton solution in detail to investigate whether the solutions include a black ring solution without conical singularity or not. We found the works\cite{Y2}\cite{E}-\cite{SR} in which the 5D EMd equations are studied and the black ring solutions are obtained, but the methods used there are different from our explicit construction of an infinite number of solutions.

The reason why we can apply the ISM to the 5D EMd equations is that there is a similarity between the 2D soliton equation and the higher-dimensional Einstein equation with axial symmetry. The higher-dimensional Einstein equation with axial symmetry is expressed in terms of the canonical variables $\rho$ and  $z$. Both the soliton and the Einstein equations are written as a 2D zero-curvature equation with these variables. When we add the dilaton  and electro-magnetic fields to the vacuum Einstein equation, the EMd system does not seem to be cast to the 2D zero-curvature equation at the first glance, but we are able to derive the 2D zero-curvature equation after introducing new functions. We already used a similar method to study equilibrium of two dilaton black holes with electric charges in four dimensions\cite{AK4D5}.  

This paper is constructed as follows. In the following section, we show an infinite number of soliton solutions to the EMd equations by introducing new functions. The new functions are fit to express them as the 2D zero-curvature equation, which enables us to yield the soliton solutions. We leave the value of the dilaton coupling strength arbitrary throughout this paper. In section 3, we discuss the 2-soliton solution in detail, which generally corresponds to black ring and/or black hole solutions. We analyse the solution by studying the rod structure of the solution, and show the 2-soliton solution is a black ring solution with electric and dilatonic charges but there is a conical singularity. By removing this conical singularity, we obtain the 5D dilatonic black hole solution. The last section is devoted to summary and discussion.

%%%%%%%%%%%%%%%%%%%%%%%%%%%%%%%%%%%%%%%%%
%
%    section 2
%
%%%%%%%%%%%%%%%%%%%%%%%%%%%%%%%%%%%%%%%%%
\section{Solutions of Einstein-Maxwell-dilaton equations}
%%%%%%%%%%%%%%%%%%%%%%%EMd equations%%%%%%%%%%%%%%%%%%%%%%%%%%%%%%%%%
The action of 5D EMd gravity is given by
\be
S=\frac{1}{16\pi}\int d^5x\sqrt{-g}
\left(R-\frac{8}{3}g^{\mu \nu}\partial_{\mu}\Phi \partial_{\nu}\Phi-\frac{1}{2}e^{-\frac{4}{3}\alpha \Phi}F_{\mu \nu}F^{\mu \nu}\right),
\ee
and metric we consider is given by
\be
ds^2=f(d\rho^2+dz^2)+g_{ab}dx^adx^b, \quad (a,b=0,1,2)
\ee
where $f$ and $g_{ab}$ are functions of $\rho$ and $z$. The 5D EMd equations read
\be
&&R^\mu{}_\nu=\frac{4}{3}\partial^\mu \Phi\partial_\nu \Phi+2e^{-\frac{4}{3}\alpha \Phi}\left( F^{\mu\sigma}F_{\nu\sigma}
-\frac{1}{6}\delta^\mu{}_\nu
F^{\beta\sigma}F_{\beta\sigma} \right),\\
&&\nabla^2\Phi=-\frac{\alpha}{2} e^{-\frac{4}{3}\alpha \Phi}F_{\beta\sigma}F^{\beta\sigma},\\
&&(e^{-\frac{4}{3}\alpha \Phi}F^{\mu\nu}{})_{;\mu}=0,
\ee
with
\be
&&F_{\mu\nu}=A_\nu,_\mu-A_\mu,_\nu.
\ee
Here $\alpha$ is a constant. We solve these equations under the static condition with the coordinate condition $\det g=-\rho^2$. We also assume that there is only electric charge. Then, the part of the metric $g=(g_{ab})$ and the $U(1)$ gauge field are assumed to have the following form:
\be
&&g={\rm diag}\left(-h_1^{-1}h_2^{-1},
\left[\sqrt{\rho^2+z^2}-z\right]h_1,
\left[\sqrt{\rho^2+z^2}+z\right]h_2\right),
\\
&&A_0=-\chi,\quad A_1=A_2=A_\rho=A_z=0,
\ee
where $h_1$ and $h_2$ are functions of $\rho$ and $z$, and $\chi(\rho,z)$ is the electrostatic potential. Then, the EMd equations are explicitly written as
\be
&&[\rho (\ln h_i),_\rho],_\rho+[\rho (\ln h_i),_z],_z
=-\frac{4\rho}{3}h_1h_2 e^{-\frac{4}{3}\alpha \Phi}(\chi,_\rho^2+\chi,_z^2), \quad (i=1,2)\label{dem1}\\
&&(\ln f),_\rho+\frac{\rho}{\rho^2+z^2}-\frac{1}{2}\Biggr(
\frac{\rho}{\sqrt{\rho^2+z^2}}[(\ln h_1),_z-(\ln h_2),_z] \cr
&&+\frac{\sqrt{\rho^2+z^2}+z}{\sqrt{\rho^2+z^2}}(\ln h_1),_\rho
+\frac{\sqrt{\rho^2+z^2}-z}{\sqrt{\rho^2+z^2}}(\ln h_2),_\rho \cr
&&+\rho[(\ln h_1),_\rho^2+(\ln h_2),_\rho^2+(\ln h_1),_\rho(\ln h_2),_\rho] \cr
&&-\rho[(\ln h_1),_z^2+(\ln h_2),_z^2+(\ln h_1),_z(\ln h_2),_z]
\Biggr)\cr
&&=-2\rho\left[h_1h_2 e^{-\frac{4}{3}\alpha \Phi}(\chi,_\rho^2-\chi,_z^2)-\frac{2}{3}(\Phi,_\rho^2-\Phi,_z^2)\right]
,\label{dem2}\\
&&(\ln f),_z+\frac{z}{\rho^2+z^2}-\frac{1}{2}\Biggr(
-\frac{\rho}{\sqrt{\rho^2+z^2}}[(\ln h_1),_\rho-(\ln h_2),_\rho] \cr
&&+\frac{\sqrt{\rho^2+z^2}+z}{\sqrt{\rho^2+z^2}}(\ln h_1),_z
+\frac{\sqrt{\rho^2+z^2}-z}{\sqrt{\rho^2+z^2}}(\ln h_2),_z \cr
&&+2\rho[(\ln h_1),_\rho (\ln h_1),_z+(\ln h_2),_\rho (\ln h_2),_z] \cr
&&+\rho[(\ln h_1),_\rho (\ln h_2),_z+(\ln h_1),_z (\ln h_2),_\rho]
\Biggr)\cr
&&=-4\rho\left(h_1h_2 e^{-\frac{4}{3}\alpha \Phi}\chi,_\rho\chi,_z-\frac{2}{3}\Phi,_\rho\Phi,_z\right)
,\label{dem3}\\
&&(\rho h_1h_2\chi,_\rho),_\rho+(\rho h_1h_2\chi,_z),_z=\frac{4\rho}{3}\alpha h_1h_2(\Phi,_\rho \chi,_\rho+\Phi,_z\chi,_z),\label{dem4}\\
&&\nabla^2\Phi=\alpha h_1 h_2 e^{-\frac{4}{3}\alpha \Phi}({\chi},_\rho^2+\chi ,_z^2),\label{dem5}
\ee
where $\nabla ^2=\partial_\rho^2+\partial_\rho/\rho+\partial_z^2$.
In order to solve these equations, we first solve Eqs.(\ref{dem1}), (\ref{dem4}) and (\ref{dem5}) to obtain $h_i$, $\Phi$ and $\chi$, and then by substituting the solutions into Eqs.(\ref{dem2}) and (\ref{dem3}), we integrate them to obtain $f$. The technique is introductions of new functions by which the equations become simpler. 

%%%%%%%%%%%%%%%%introduction of H_i and \bar H_i%%%%%%%%%%%%%%%%%%%%%%
Note that Eq.~(\ref{dem1}) is written as
\be
\nabla^2 (\ln h_i)=-\frac{4}{3}h_1h_2 e^{-\frac{4}{3}\alpha \Phi}(\chi,_\rho^2+\chi,_z^2).\quad (i=1,2)\label{dem1a}
\ee
Then, combining these equations with Eq.~(\ref{dem5}), we obtain
\be
\nabla^2 (\ln h_i+\frac{2\alpha}{3} \Phi)&=&\frac{2}{3}(\alpha^2-2)h_1h_2 e^{-\frac{4}{3}\alpha \Phi}(\chi,_\rho^2+\chi,_z^2),\quad (i=1,2)\\
\nabla^2 (\ln h_i-\frac{2\alpha}{3} \Phi)&=&-\frac{2}{3}(\alpha^2+2)h_1h_2 e^{-\frac{4}{3}\alpha \Phi}(\chi,_\rho^2+\chi,_z^2).\quad (i=1,2)
\ee
By defining $H_i$ and $\bar H_i$ by
\be
H_i=h_ie^{\frac{2\alpha}{3} \Phi},~\bar H_i=h_ie^{-\frac{2\alpha}{3} \Phi},~(i=1,2)\label{defofH}
\ee
these equations are rewritten as
\be
\nabla^2 \ln H_i&=&\frac{2}{3}(\alpha^2-2)\bar H_1 \bar H_2 (\chi,_\rho^2+\chi,_z^2),\quad (i=1,2)\label{eqH}\\
\nabla^2 \ln \bar H_i&=&-\frac{2}{3}(\alpha^2+2) \bar H_1 \bar H_2 (\chi,_\rho^2+\chi,_z^2),\quad (i=1,2)\label{eqbarH}
\ee
and Eq.~(\ref{dem4}) reads
\be
(\rho \bar H_1\bar H_2\chi,_\rho),_\rho+
(\rho \bar H_1\bar H_2\chi,_z),_z=0. \label{eqchi}
\ee
From Eq.~(\ref{defofH}) we have the relations
\be
\frac{H_1}{H_2}=\frac{\bar H_1}{\bar H_2}=\frac{h_1}{h_2}, \label{relamgH}
\ee
and the equations in Eq.~(\ref{defofH}) are inversely solved to yield the relations
\be
\Phi&=&\frac{3}{4\alpha}(\ln H_i-\ln \bar H_i),\quad (i=1,2)\label{Phi}\\
\ln h_i&=&\frac{1}{2}(\ln H_i+\ln \bar H_i).\quad (i=1,2)\label{lnh}
\ee
By using these expressions, Eqs.~(\ref{dem2}) and (\ref{dem3}) are rewritten as
\be
&&(\ln f),_\rho+\frac{\rho}{\rho^2+z^2}-\frac{1}{4}\Biggr(
\frac{\rho}{\sqrt{\rho^2+z^2}}[(\ln H_1),_z-(\ln H_2),_z] \cr
&&+\frac{\sqrt{\rho^2+z^2}+z}{\sqrt{\rho^2+z^2}}(\ln H_1),_\rho
+\frac{\sqrt{\rho^2+z^2}-z}{\sqrt{\rho^2+z^2}}(\ln H_2),_\rho \cr
&&+\frac{\alpha^2+2}{2 \alpha^2} \rho \left[(\ln H_1),_\rho^2+(\ln H_2),_\rho^2+(\ln H_1),_\rho(\ln H_2),_\rho \right.\cr
&&\left. -(\ln H_1),_z^2-(\ln H_2),_z^2-(\ln H_1),_z(\ln H_2),_z\right]
\Biggr) \cr
&&-\frac{1}{4}\Biggr(
\frac{\rho}{\sqrt{\rho^2+z^2}}[(\ln \bar H_1),_z-(\ln \bar H_2),_z] \cr
&&+\frac{\sqrt{\rho^2+z^2}+z}{\sqrt{\rho^2+z^2}}(\ln \bar H_1),_\rho
+\frac{\sqrt{\rho^2+z^2}-z}{\sqrt{\rho^2+z^2}}(\ln \bar H_2),_\rho \cr
&&+\frac{\alpha^2+2}{2 \alpha^2} \rho\left[ (\ln \bar H_1),_\rho^2+(\ln \bar H_2),_\rho^2+(\ln \bar H_1),_\rho(\ln \bar H_2),_\rho \right.\cr
&&\left.-(\ln \bar H_1),_z^2-(\ln \bar H_2),_z^2-(\ln \bar H_1),_z(\ln \bar H_2),_z\right] \Biggr) \cr
&&-\frac{\alpha^2-2}{8 \alpha^2} \rho 
\left[2(\ln H_1),_\rho (\ln \bar H_1),_\rho+2(\ln H_2),_\rho (\ln \bar H_2),_\rho \right.\cr
&&+(\ln H_1),_\rho (\ln \bar H_2),_\rho+(\ln H_2),_\rho (\ln \bar H_1),_\rho \cr
&&-2(\ln H_1),_z (\ln \bar H_1),_z-2(\ln H_2),_z (\ln \bar H_2),_z \cr
&&\left.-(\ln H_1),_z (\ln \bar H_2),_z-(\ln H_2),_z (\ln \bar H_1),_z \right]
=-2\rho \bar H_1\bar H_2 (\chi,_\rho^2-\chi,_z^2)
,\label{dem2a}\\
&&(\ln f),_z+\frac{z}{\rho^2+z^2}-\frac{1}{4}\Biggr(
-\frac{\rho}{\sqrt{\rho^2+z^2}}[(\ln H_1),_\rho-(\ln H_2),_\rho] \cr
&&+\frac{\sqrt{\rho^2+z^2}+z}{\sqrt{\rho^2+z^2}}(\ln H_1),_z
+\frac{\sqrt{\rho^2+z^2}-z}{\sqrt{\rho^2+z^2}}(\ln H_2),_z \cr
&&+\frac{\alpha^2+2}{2\alpha^2}\rho \left[2(\ln H_1),_\rho (\ln H_1),_z
+2(\ln H_2),_\rho (\ln H_2),_z \right.\cr
&&\left. +(\ln H_1),_\rho (\ln H_2),_z
+(\ln H_1),_z (\ln H_2),_\rho \right] \Biggr) \cr
&&-\frac{1}{4}\Biggr(
-\frac{\rho}{\sqrt{\rho^2+z^2}}[(\ln \bar H_1),_\rho-(\ln \bar H_2),_\rho] \cr
&&+\frac{\sqrt{\rho^2+z^2}+z}{\sqrt{\rho^2+z^2}}(\ln \bar H_1),_z
+\frac{\sqrt{\rho^2+z^2}-z}{\sqrt{\rho^2+z^2}}(\ln \bar H_2),_z \cr
&&+\frac{\alpha^2+2}{2\alpha^2}\rho\left[2(\ln \bar H_1),_\rho (\ln \bar H_1),_z+2(\ln \bar H_2),_\rho (\ln \bar H_2),_z \right. \cr
&&\left.+(\ln \bar H_1),_\rho (\ln \bar H_2),_z+(\ln \bar H_1),_z (\ln \bar H_2),_\rho \right] \Biggr) \cr
&&-\frac{\alpha^2-2}{8 \alpha^2}\rho \left[ 2(\ln H_1),_\rho (\ln \bar H_1),_z+2(\ln H_2),_\rho(\ln \bar H_2),_z \right.\cr
&&+2(\ln \bar H_1),_\rho (\ln H_1),_z 
+2(\ln \bar H_2),_\rho(\ln H_2),_z+
(\ln H_1),_\rho (\ln \bar H_2),_z \cr
&&\left.+(\ln H_2),_\rho(\ln \bar H_1),_z 
+(\ln \bar H_1),_\rho (\ln  H_2),_z+(\ln \bar H_2),_\rho(\ln H_1),_z \right] \cr
&&=-4\rho \bar H_1 \bar H_2 \chi,_\rho \chi,_z
,\label{dem3a}
\ee

%%%%%%%%%%%%%%%%%relation between H_i and \bar H_i%%%%%%%%%%%%%%%%%%%
From Eqs.~(\ref{eqH}) and (\ref{eqbarH}) we find $H_i$'s and $\bar H_i$'s are related up to harmonic functions. Therefore, they are related to each other as
\be
\ln H_i=-\frac{\alpha^2-2}{\alpha^2+2}\ln\bar H_i+\ln N_i,\quad (i=1,2)
\label{relH}
\ee
with the equation
\be
\nabla^2 \ln N_i=0.\quad (i=1,2)\label{eqNi}
\ee
Substituting Eq.~(\ref{relH}) into Eqs.~(\ref{dem2a}) and (\ref{dem3a}), we have
\be
&&(\ln f),_\rho+\frac{\rho}{\rho^2+z^2}-\frac{1}{\alpha^2+2}\Biggr(
\frac{\rho}{\sqrt{\rho^2+z^2}}[(\ln \bar H_1),_\rho-(\ln \bar H_2),_\rho] \cr
&&+\frac{\sqrt{\rho^2+z^2}+z}{\sqrt{\rho^2+z^2}}(\ln \bar H_1),_z
+\frac{\sqrt{\rho^2+z^2}-z}{\sqrt{\rho^2+z^2}}(\ln \bar H_2),_z \cr
&&+\rho\left[ (\ln \bar H_1),_\rho^2+(\ln \bar H_2),_\rho^2
+(\ln \bar H_1),_\rho(\ln \bar H_2),_\rho \right.\cr
&&\left.-(\ln \bar H_1),_z^2-(\ln \bar H_2),_z^2-(\ln \bar H_1),_z(\ln \bar H_2),_z\right] \Biggr) \cr
&&-\frac{1}{4}\Biggr(
\frac{\rho}{\sqrt{\rho^2+z^2}}[(\ln N_1),_z-(\ln N_2),_z] \cr
&&+\frac{\sqrt{\rho^2+z^2}+z}{\sqrt{\rho^2+z^2}}(\ln N_1),_\rho
+\frac{\sqrt{\rho^2+z^2}-z}{\sqrt{\rho^2+z^2}}(\ln N_2),_\rho \cr
&&+\frac{\alpha^2+2}{2 \alpha^2} \rho \left[(\ln N_1),_\rho^2+(\ln N_2),_\rho^2+(\ln N_1),_\rho(\ln N_2),_\rho \right.\cr
&&\left. -(\ln N_1),_z^2-(\ln N_2),_z^2-(\ln N_1),_z(\ln N_2),_z\right]
\Biggr) \cr
&&=-2\rho \bar H_1\bar H_2 (\chi,_\rho^2-\chi,_z^2),\label{lnfrho}\\
&&(\ln f),_z+\frac{z}{\rho^2+z^2}
-\frac{1}{\alpha^2+2}\Biggr(
-\frac{\rho}{\sqrt{\rho^2+z^2}}[(\ln \bar H_1),_\rho-(\ln \bar H_2),_\rho] \cr
&&+\frac{\sqrt{\rho^2+z^2}+z}{\sqrt{\rho^2+z^2}}(\ln \bar H_1),_z
+\frac{\sqrt{\rho^2+z^2}-z}{\sqrt{\rho^2+z^2}}(\ln \bar H_2),_z \cr
&&+\rho\left[2(\ln \bar H_1),_\rho (\ln \bar H_1),_z
+2(\ln \bar H_2),_\rho (\ln \bar H_2),_z \right. \cr
&&\left.+(\ln \bar H_1),_\rho (\ln \bar H_2),_z
+(\ln \bar H_1),_z (\ln \bar H_2),_\rho \right] \Biggr) \cr
&&-\frac{1}{4}\Biggr(
-\frac{\rho}{\sqrt{\rho^2+z^2}}[(\ln N_1),_\rho-(\ln N_2),_\rho] \cr
&&+\frac{\sqrt{\rho^2+z^2}+z}{\sqrt{\rho^2+z^2}}(\ln N_1),_z
+\frac{\sqrt{\rho^2+z^2}-z}{\sqrt{\rho^2+z^2}}(\ln N_2),_z \cr
&&+\frac{\alpha^2+2}{2\alpha^2}\rho \left[2(\ln N_1),_\rho (\ln N_1),_z
+2(\ln N_2),_\rho (\ln N_2),_z \right.\cr
&&\left. +(\ln N_1),_\rho (\ln N_2),_z
+(\ln N_1),_z (\ln N_2),_\rho \right] \Biggr)
=-4\rho \bar H_1 \bar H_2 \chi,_\rho \chi,_z.
\label{lnfz}
\ee
In order to solve these equations, we introduce $f_N$ and $f_{em}$ by
\be
f=\sqrt{f_N\cdot f_{em}}, \label{fNfem}
\ee
and require that $f_N$ should satisfy
\be
&&(\ln f_N),_\rho+\frac{\rho}{\rho^2+z^2}-\frac{1}{2}\Biggr(
\frac{\rho}{\sqrt{\rho^2+z^2}}[(\ln N_1),_z-(\ln N_2),_z] \cr
&&+\frac{\sqrt{\rho^2+z^2}+z}{\sqrt{\rho^2+z^2}}(\ln N_1),_\rho
+\frac{\sqrt{\rho^2+z^2}-z}{\sqrt{\rho^2+z^2}}(\ln N_2),_\rho \cr
&&+\frac{\alpha^2+2}{2 \alpha^2} \rho \left[(\ln N_1),_\rho^2+(\ln N_2),_\rho^2+(\ln N_1),_\rho(\ln N_2),_\rho \right.\cr
&&\left. -(\ln N_1),_z^2-(\ln N_2),_z^2-(\ln N_1),_z(\ln N_2),_z\right]
\Biggr)=0, \label{eqfNrho}\\
&&(\ln f_N),_z+\frac{z}{\rho^2+z^2}-\frac{1}{2}\Biggr(
-\frac{\rho}{\sqrt{\rho^2+z^2}}[(\ln N_1),_\rho-(\ln N_2),_\rho] \cr
&&+\frac{\sqrt{\rho^2+z^2}+z}{\sqrt{\rho^2+z^2}}(\ln N_1),_z
+\frac{\sqrt{\rho^2+z^2}-z}{\sqrt{\rho^2+z^2}}(\ln N_2),_z \cr
&&+\frac{\alpha^2+2}{2\alpha^2}\rho \left[2(\ln N_1),_\rho (\ln N_1),_z
+2(\ln N_2),_\rho (\ln N_2),_z \right.\cr
&&\left. +(\ln N_1),_\rho (\ln N_2),_z
+(\ln N_1),_z (\ln N_2),_\rho \right]
\Biggr)=0,\label{eqfNz}
\ee
which have similar forms to those of the corresponding equations in the 5D static vacuum case. Then, $f_{em}$ should satisfy
\be
&&(\ln f_{em}),_\rho+\frac{\rho}{\rho^2+z^2}-\frac{2}{\alpha^2+2}\Biggr(
\frac{\rho}{\sqrt{\rho^2+z^2}}[(\ln \bar H_1),_\rho-(\ln \bar H_2),_\rho] \cr
&&+\frac{\sqrt{\rho^2+z^2}+z}{\sqrt{\rho^2+z^2}}(\ln \bar H_1),_z
+\frac{\sqrt{\rho^2+z^2}-z}{\sqrt{\rho^2+z^2}}(\ln \bar H_2),_z \cr
&&+\rho\left[ (\ln \bar H_1),_\rho^2+(\ln \bar H_2),_\rho^2
+(\ln \bar H_1),_\rho(\ln \bar H_2),_\rho \right.\cr
&&\left.-(\ln \bar H_1),_z^2-(\ln \bar H_2),_z^2-(\ln \bar H_1),_z(\ln \bar H_2),_z\right] \Biggr) \cr
&&=-4\rho \bar H_1\bar H_2 (\chi,_\rho^2-\chi,_z^2),
\label{eqfemrho}\\
&&(\ln f_{em}),_z+\frac{z}{\rho^2+z^2}-\frac{2}{\alpha^2+2}\Biggr(
-\frac{\rho}{\sqrt{\rho^2+z^2}}[(\ln \bar H_1),_\rho-(\ln \bar H_2),_\rho] \cr
&&+\frac{\sqrt{\rho^2+z^2}+z}{\sqrt{\rho^2+z^2}}(\ln \bar H_1),_z
+\frac{\sqrt{\rho^2+z^2}-z}{\sqrt{\rho^2+z^2}}(\ln \bar H_2),_z \cr
&&+\rho\left[2(\ln \bar H_1),_\rho (\ln \bar H_1),_z
+2(\ln \bar H_2),_\rho (\ln \bar H_2),_z \right. \cr
&&\left.+(\ln \bar H_1),_\rho (\ln \bar H_2),_z
+(\ln \bar H_1),_z (\ln \bar H_2),_\rho \right] \Biggr)
=-8\rho \bar H_1 \bar H_2 \chi,_\rho \chi,_z,
\label{eqfemz}
\ee
which have similar forms to those of the corresponding equations in the 5D electro-static case.

%%%%%%%%%%%%%%%%%%%%forms of \bar H_i%%%%%%%%%%%%%%%%%%%%%%%%%%
In order to solve Eqs.~(\ref{eqbarH}) and (\ref{eqchi}), we assume that the functions $\bar H_1$ and $\bar H_2$ have following forms
\be
\bar H_1&=&\left[ 1-\frac{4}{3}(\alpha^2+2)c\chi+\frac{2}{3}(\alpha^2+2)\chi^2\right]^{-1/2}N^{1/2},\label{barH1}\\
\bar H_2&=&\left[ 1-\frac{4}{3}(\alpha^2+2)c\chi+\frac{2}{3}(\alpha^2+2)\chi^2\right]^{-1/2}N^{-1/2},\label{barH2}
\ee
where $N$ is a function of $\rho$ and $z$, and $c$ is a constant. Then,  Eqs.~(\ref{eqbarH}) and (\ref{eqchi}) are both put into the form
\be
\nabla^2\chi=\frac{4(\alpha^2+2)(\chi-c)}
{3-4(\alpha^2+2)c\chi+2(\alpha^2+2)\chi^2}(\chi,_\rho^2+\chi,_z^2),
\label{chi}
\ee
with the Laplace equation for $\ln N$
\be
\nabla^2\ln N=0.\label{lap1}
\ee
Next, we introduce a new function $R(\rho, z)$ through the relation
\be
\chi=\frac{e}{R+m},
\ee 
where $e$ and $m$ are constants satisfying
\be
m=\frac{2(\alpha^2+2)}{3}ce.
\ee
Then, Eqs.~(\ref{barH1}) and (\ref{barH2}) lead to
\be
\bar H_1&=&\frac{m+R}{\sqrt{R^2-d^2}}N^{1/2},\label{barH1a}\\
\bar H_2&=&\frac{m+R}{\sqrt{R^2-d^2}}N^{-1/2},\label{barH2a}
\ee
where
\be
d=\sqrt{\frac{3m^2-2(\alpha^2+2)e^2}{3}}.
\ee
Equation (\ref{chi}) can also be expressed in terms of $R$ as
\be
\nabla^2R=2R(R^2-d^2)^{-1}(R,_\rho^2+R,_z^2).
\ee
Then, introducing a function $h(\rho, z)$ through the relation
\be
R=d\frac{1+h}{1-h},\label{Rh}
\ee
we obtain the Laplace equation for $\ln h$ as
\be
\nabla^2\ln h=0. \label{lap2}
\ee
Note that we can construct the solution to Eq.(\ref{chi}) for the electric potential $\chi$ via $R$ given in Eq.(\ref{Rh}).
We rewrite  (\ref{eqfemrho}) and (\ref{eqfemz}) in terms of $\chi$ and $N$.
Equation (\ref{eqfemrho}) is written as
\be
&&(\ln f_{em}),_\rho=-\frac{\rho}{\rho^2+z^2}
-\frac{8(\chi-c)\chi,_\rho}{3-4(\alpha^2+2)c\chi+2(\alpha^2+2)\chi^2} \cr
&&+\frac{12\rho[2(\alpha^2+2)c^2-3)](\chi,_\rho^2-\chi,_z^2)}
{[3-4(\alpha^2+2)c\chi+2(\alpha^2+2)\chi^2]^2} \cr
&&+\frac{2}{\alpha^2+2}\left(
\frac{\left[z(\ln N),_\rho+\rho(\ln N),_z\right]}
{\sqrt{\rho^2+z^2}}
+\frac{\rho}{4}\left[(\ln N),_\rho^2-(\ln N),_z^2\right]\right),
\label{femrho}
\ee
and Eq.~(\ref{eqfemz}) as
\be
&&(\ln f_{em}),_z=-\frac{z}{\rho^2+z^2}
-\frac{8(\chi-c)\chi,_z}{3-4(\alpha^2+2)c\chi+2(\alpha^2+2)\chi^2} \cr
&&+\frac{24\rho[2(\alpha^2+2)c^2-3)]\chi,_\rho\chi,_z}
{[3-4(\alpha^2+2)c\chi+2(\alpha^2+2)\chi^2]^2} \cr
&&+\frac{2}{\alpha^2+2}\left(
-\frac{\left[\rho(\ln N),_\rho-z(\ln N),_z\right]}{\sqrt{\rho^2+z^2}}
+\frac{\rho}{2}(\ln N),_\rho(\ln N),_z\right).\label{femz}
\ee
By further introducing $Q_{em}(\rho, z)$ through
\be
f_{em}=\frac{1}{2\sqrt{\rho^2+z^2}}
\left[ 1-\frac{4}{3}(\alpha^2+2)c\chi+\frac{2}{3}(\alpha^2+2)\chi^2\right]^{-2/(\alpha^2+2)}Q_{em},\label{defQem}
\ee
Eqs.~(\ref{femrho}) and (\ref{femz}) are expressed as
\be
(\ln Q_{em}),_\rho=\frac{2}{\alpha^2+2}&\Biggr(&
\frac{3\rho}{4}\left[(\ln h),_\rho^2-(\ln h),_z^2\right] 
+\frac{\rho}{4}\left[(\ln N),_\rho^2-(\ln N),_z^2\right] \cr
&&+\frac{\left[z(\ln N),_\rho+\rho(\ln N),_z\right]}{\sqrt{\rho^2+z^2}}\Biggr)
\label{Qemrho}
\ee
and
\be
(\ln Q_{em}),_z=\frac{2}{\alpha^2+2}&\Biggr(&
\frac{3\rho}{2}(\ln h),_\rho(\ln h),_z
+\frac{\rho}{2}(\ln N),_\rho(\ln N),_z \cr
&&-\frac{\left[\rho(\ln N),_\rho-z(\ln N),_z\right]}{\sqrt{\rho^2+z^2}}
\Biggr).\label{Qemz}
\ee
We next rewrite Eqs.~(\ref{eqfNrho}) and (\ref{eqfNz}) in terms of $N$ and $N_2$. We rewrite these equations by introducing $Q_N(\rho,z)$ through
\be
f_N=\frac{1}{2\sqrt{\rho^2+z^2}}Q_N,\label{defQN}
\ee
and taking account of the relation
\be
\frac{N_1}{N_2}=N^{2\alpha^2/(\alpha^2+2)},\label{ratio}
\ee
Eqs.~(\ref{eqfNrho}) and (\ref{eqfNz}) are reduced to
\be
&&(\ln Q_N),_\rho=(\ln N_2),_\rho
+\frac{3(\alpha^2+2)\rho}{4\alpha^2}
\left[(\ln N_2),_\rho^2-(\ln N_2),_z^2\right]\cr
&&+\frac{3\rho}{2}
\left[(\ln N_2),_\rho(\ln N),_\rho-(\ln N_2),_z(\ln N),_z\right]\cr
&&+\frac{\alpha^2}{\alpha^2+2}\Biggr((\ln N),_\rho
+\frac{z(\ln N),_\rho+\rho(\ln N),_z}{\sqrt{\rho^2+z^2}}
+\rho\left[(\ln N),_\rho^2-(\ln N),_z^2\right]\Biggr), \quad\quad
\label{Q0rho} \\
&&(\ln Q_N),_z=(\ln N_2),_z
+\frac{3(\alpha^2+2)\rho}{2\alpha^2}(\ln N_2),_\rho(\ln N_2),_z \cr
&&+\frac{3\rho}{2}
\left[(\ln N_2),_\rho(\ln N),_z+(\ln N_2),_z(\ln N),_z\right] \cr
&&+\frac{\alpha^2}{\alpha^2+2}\Biggr((\ln N),_z
-\frac{\left[\rho(\ln N),_\rho-z(\ln N),_z\right]}{\sqrt{\rho^2+z^2}}
+2\rho(\ln N),_\rho(\ln N),_z\Biggr). \quad\quad
\label{Q0z}
\ee

%%%%%%%%%%%%relations among \ln h, \ln N and \ln N_2%%%%%%%%%%%%%%%%%%
As is seen from Eqs.~(\ref{eqNi}), (\ref{lap1}) and (\ref{lap2}), $\ln N_2$, $\ln N$ and $\ln h$ all satisfy the Laplace equation. When we restrict ourselves only to the soliton solutions, they can have solutions with different soliton numbers in general. However, they are too rich in analysing them. In order to clarify the structure of the solutions, we simplify the solution by assuming the ansatz that $\ln N_2$ and $\ln N$ can be expressed in terms of $\ln h$:
\be
\ln N_2&=&\frac{2\alpha^2k_1}{\alpha^2+2} \ln h,\label{ansatz1}\\
\ln N&=&k_2 \ln h,\label{ansatz2}
\ee
where $k_1$ and $k_2$ are constants. Then, the functions $h_1$ and $h_2$ in Eq.~(\ref{lnh}) are written in terms of $h$ through the relations (\ref{relH}), (\ref{barH1a}), (\ref{barH2a}) and (\ref{ratio}) as
\be
h_1&=&\left[\frac{d(1+h)+m(1-h)}{2d}\right]^{2/(\alpha^2+2)}
h^{[\alpha^2k_1+(\alpha^2+1)k_2-1]/(\alpha^2+2)},\\
h_2&=&\left[\frac{d(1+h)+m(1-h)}{2d}\right]^{2/(\alpha^2+2)}
h^{(\alpha^2k_1-k_2-1)/(\alpha^2+2)}.
\ee
From the definitions (\ref{fNfem}), (\ref{defQem}) and (\ref{defQN}), we find that $f$ is now expressed as
\be
f=\frac{1}{2\sqrt{\rho^2+z^2}}
\left[ 1-\frac{4}{3}(\alpha^2+2)c\chi+\frac{2}{3}(\alpha^2+2)\chi^2\right]^{-1/(\alpha^2+2)}\sqrt{Q_N Q_{em}}.
\ee 
The quantity $Q$ defined by $Q=\sqrt{Q_NQ_{em}}$ has a simple expression by using the above ansatz as
\be
(\ln Q),_\rho&=&\frac{\alpha^2(2k_1+k_2)}{2(\alpha^2+2)}(\ln h),_\rho+
\frac{\rho}{2}\nu[(\ln h),_\rho^2-(\ln h),_z^2] \cr
&&+\frac{k_2}{2\sqrt{\rho^2+z^2}}
\left[z(\ln h),_\rho+\rho(\ln h),_z\right], \label{Qrho}\\
(\ln Q),_z&=&\frac{\alpha^2(2k_1+k_2)}{2(\alpha^2+2)}(\ln h),_z+
\rho \nu(\ln h),_\rho(\ln h),_z \cr
&&-\frac{k_2}{2\sqrt{\rho^2+z^2}}
\left[\rho(\ln h),_\rho-z(\ln h),_z\right],
\label{Qz}
\ee
where
\be
\nu&=&\frac{1}{\alpha^2+2}\left(\frac{3}{2}+3\alpha^2k_1^2+3\alpha^2k_1k_2+\frac{2\alpha^2+1}{2}k_2^2 \right).
\ee

%%%%%%%%%%%%%%%%%%%n-soliton solution%%%%%%%%%%%%%%%%%%%%%%%%%%%%%%
The $n$-soliton solution for $h$ is given by
\be
h=\frac{\prod_k^n(i\mu_k)}{\rho^n}, \label{n_soliton}
\ee
with the pole trajectories
\be
\mu_k=w_k-z+(-1)^{k-1}\sqrt{(w_k-z)^2+\rho^2}, \quad (k=1,2,\cdots,n) 
\ee
where $w_k(k=1,2,\cdots,n)$ are constants.
By using the formulas
\be
\frac{\rho}{2}[(\ln h),_\rho^2-(\ln h),_z^2]&=&\frac{\partial}{\partial \rho}\ln \left[\frac{\rho^{n^2/2}\prod(\mu_k-\mu_l)^2}{\prod(i\mu_k)^{n-2}\prod(\mu_k^2+\rho^2)}\right],\\
\frac{[z (\ln h),_{\rho}+\rho(\ln h),_z]}{2\sqrt{\rho^2+z^2}}
&=&\frac{\partial}{\partial \rho}\ln 
\left[\frac{(\sqrt{\rho^2+z^2}+z)^{n/2}\prod(i\mu_k)^{1/2}}{\rho^{n/2}\prod(\sqrt{\rho^2+z^2}+z+\mu_k)}\right],
\ee
we obtain
\be
Q=C^{(n)}\frac{\rho^{\nu_1}\prod_k^n(i\mu_k)^{\nu_2}\prod_{k>l}^n(\mu_k-\mu_l)^{2\nu}(\sqrt{\rho^2+z^2}+z)^{nk_2/2}}{\prod_k^n(\sqrt{\rho^2+z^2}+z+\mu_k)^{k_2}\prod_k^n(\mu_k^2+\rho^2)^{\nu}},
\ee
where $C^{(n)}$ is a constant to be determined by the asymptotic flatness condition and
\be
\nu_1&=&\frac{n^2}{2}\nu-\frac{n}{\alpha^2+2}[\alpha^2k_1+(\alpha^2+1)k_2],\\
\nu_2&=&-(n-2)\nu+\frac{1}{\alpha^2+2}[\alpha^2k_1+(\alpha^2+1)k_2].
\ee
Here we summarize the result. The metric components are given by
\be
g_{00}&=&-K^{-1}
h^{-(2\alpha^2k_1+\alpha^2k_2-2)/(\alpha^2+2)},\label{nsolitong00}\\
g_{11}&=&(\sqrt{\rho^2+z^2}-z)\sqrt{K} h^{[\alpha^2k_1+(\alpha^2+1)k_2-1]/(\alpha^2+2)},\\
g_{22}&=&-(\sqrt{\rho^2+z^2}+z)\sqrt{K} h^{(\alpha^2k_1-k_2-1)/(\alpha^2+2)},\\
g_{\rho \rho}&=&g_{zz}=\frac{C^{(n)}}{2\sqrt{\rho^2+z^2}}
\sqrt{K} \cr
&{}&\qquad\times \ \
\frac{\rho^{\nu_1}\prod_k^n(i\mu_k)^{\nu_2}\prod_{k>l}^n(\mu_k-\mu_l)^{2\nu}(\sqrt{\rho^2+z^2}+z)^{nk_2/2}}{\prod_k^n(\sqrt{\rho^2+z^2}+z+\mu_k)^{k_2}\prod_k^n(\mu_k^2+\rho^2)^{\nu}},
\ee
and the dilaton field and the electric potential are given by
\be
\Phi&=&-\frac{3\alpha}{8}\ln K+\frac{3\alpha(1+2k_1+k_2)}{4(\alpha^2+2)}\ln h,\\
\chi&=&\frac{e(1-h)}{2d}K^{-(\alpha^2+2)/4},
\ee
where $K$ is given by
\be
K=\left[\frac{d(1+h)+m(1-h)}{2d}\right]^{4/(\alpha^2+2)},
\ee
and $h$ is given by Eq.(\ref{n_soliton}).

%%%%%%%%%%%%%%%%%%%2-soliton solution%%%%%%%%%%%%%%%%%%%%
We illustrate the solution by taking $n=2$ or 2-soliton solution case. In this case, $h$ and $Q$ are given by
\be
h&=&-\frac{\mu_1\mu_2}{\rho^2},\label{2solitonh}\\
Q&=&\left[\frac{2(z_0+d)(\sqrt{\rho^2+z^2}+z)}
{(\sqrt{\rho^2+z^2}+z+\mu_1)(\sqrt{\rho^2+z^2}+z+\mu_2)}\right]^{k_2} \cr
&\times&\left(-\frac{\mu_1\mu_2}{\rho^2}
\right)^{[\alpha^2k_1+(\alpha^2+1)k_2)]/(\alpha^2+2)}
\left[\frac{\rho^2(\mu_2-\mu_1)^2}
{(\mu_1^2+\rho^2)(\mu_2^2+\rho^2)}\right]^{\nu},
\ee
where we have put $w_1=z_0-d$ and $w_2=z_0+d$, and so $\mu_1$ and $\mu_2$ are given by
\be
\left\{
\matrix{
\mu_1=z_0-z-d+\sqrt{(z_0-z-d)^2+\rho^2},\cr
\mu_2=z_0-z+d-\sqrt{(z_0-z+d)^2+\rho^2}.
}
\right.
\ee
By transforming the variable $z$ to $z+z_0$, we rewrite these poles as
\be
\left\{
\matrix{
\mu_1=-(z+d)+\sqrt{(z+d)^2+\rho^2},\cr
\mu_2=-(z-d)-\sqrt{(z-d)^2+\rho^2},}
\right.\label{mus}
\ee
and define new poles by
\be
\left\{
\matrix{
\mu=-(z+z_0)-\sqrt{(z+z_0)^2+\rho^2},\cr
\mu^*=-(z+z_0)+\sqrt{(z+z_0)^2+\rho^2}.}
\right.
\ee
With these expressions, the 2-soliton solution is given by
\be
g_{00}&=&-K^{-1}
h^{-(2\alpha^2k_1+\alpha^2k_2-2)/(\alpha^2+2)},\label{2solitong00}\\
g_{11}&=&\mu^*\sqrt{K}
h^{[\alpha^2k_1+(\alpha^2+1)k_2-1]/(\alpha^2+2)},\\
g_{22}&=&-\mu\sqrt{K}
h^{(\alpha^2k_1-k_2-1)/(\alpha^2+2)},\\
f&=&\frac{h^{[\alpha^2k_1+(\alpha^2+1)k_2-1]/(\alpha^2+2)}}
{2\sqrt{(z+z_0)^2+\rho^2}}
\sqrt{K}\cr
&\times&\left[\frac{-2(z_0+d)\mu}{(\mu_1-\mu)(\mu_2-\mu)}\right]^{k_2}
\left[\frac{\rho^2(\mu_2-\mu_1)^2}
{(\mu_1^2+\rho^2)(\mu_2^2+\rho^2)}\right]^{\nu}.\label{2soliton}
\ee
We note that $\mu$ and $\mu^*$ have the same forms as the poles $\mu_1$ and $\mu_2$. Therefore, by regarding them as poles at different positions,  we note the metric components are completely expressed in terms of these poles. In the following section, we analyse the rod structure of this 2-soliton solution. By requiring the horizon to exist, we determine the parameters $k_1$ and $k_2$.

%%%%%%%%%%%%%%%%%%%%%%%%%%%%%%%%%%%%%%%%%
%
%    section 3
%
%%%%%%%%%%%%%%%%%%%%%%%%%%%%%%%%%%%%%%%%%
\section{Analysis of Solutions}
%%%%%%%%%%%%%%%%%%%%%%%%%rod structure%%%%%%%%%%%%%%%%%%%%%%%%%%%%
We now study $\rho\sim 0$ behaviors of the metric components, or the rod structure of the 2-soliton solution obtained in the previous section.  We divide the $z$-axis into four intervals separated by values $z=-z_0$ and $z=\pm d$ and assume that these parameters are ordered as follows, without loss of generality:
\be
-\infty<-d<d<-z_0<\infty.
\ee
If $g_{ii} \propto \rho^2~(i=1,2)$ in some interval, the rod is space-like and in $\partial/\partial x^i$ direction, and $\rho$-$i$ plane is perpendicular to the $z$-axis. Going around the $z$-axis in the $i$-th direction, we study the period of the coordinate $x^i$. The behavior of the metric components for $\rho\sim 0$ in each region is as follows.\\
\noindent
(i) $\infty>z>-z_0$
\be
g_{00}&\sim&-\left(\frac{z+m}{z+d}\right)^{-4/(\alpha^2+2)}
\left(\frac{z-d}{z+d}\right)^{-(2\alpha^2k_1+\alpha^2k_2-2)/(\alpha^2+2)},\\
g_{11}&\sim&\frac{\rho^2}{2(z+z_0)}
\left(\frac{z+m}{z+d}\right)^{2/(\alpha^2+2)}
\left(\frac{z-d}{z+d}
\right)^{[\alpha^2k_1+(\alpha^2+1)k_2-1]/(\alpha^2+2)},\\
g_{22}&\sim&2(z+z_0)\left(\frac{z+m}{z+d}\right)^{2/(\alpha^2+2)}
\left(\frac{z-d}{z+d}\right)^{(\alpha^2k_1-k_2-1)/(\alpha^2+2)},\\
f&\sim&\frac{1}{2(z+z_0)}\left(\frac{z+m}{z+d}\right)^{2/(\alpha^2+2)}
\left(\frac{z-d}{z+d}\right)^{[\alpha^2k_1+(\alpha^2+1)k_2-1]/(\alpha^2+2)}.
\ee
In this interval, we note that $g_{11}\sim {\cal O}(\rho^2)$. Since the rod is in $\partial/\partial x^1$ direction, we study the period $\triangle x^1$ to find that
\be
\triangle x^1=2\pi \lim_{\rho \to 0}\sqrt{\frac{\rho^2f}{g_{11}}}&=2\pi,
\ee
which shows no conical singularity in this interval.\\
\noindent
(ii) $-z_0>z>d$
\be
g_{00}&\sim&-\left(\frac{z+m}{z+d}\right)^{-4/(\alpha^2+2)}
\left(\frac{z-d}{z+d}\right)^{-(2\alpha^2k_1+\alpha^2k_2-2)/(\alpha^2+2)},\\
g_{11}&\sim&-2(z+z_0)\left(\frac{z+m}{z+d}\right)^{2/(\alpha^2+2)}
\left(\frac{z-d}{z+d}
\right)^{[\alpha^2k_1+(\alpha^2+1)k_2-1]/(\alpha^2+2)},\\
g_{22}&\sim&-\frac{\rho^2}{2(z+z_0)}
\left(\frac{z+m}{z+d}\right)^{2/(\alpha^2+2)}
\left(\frac{z-d}{z+d}\right)^{(\alpha^2k_1-k_2-1)/(\alpha^2+2)},\\
f&\sim&-\frac{1}{2(z+z_0)}\left(\frac{z+m}{z+d}\right)^{2/(\alpha^2+2)}
\left(\frac{z-d}{z+d}\right)^{[\alpha^2k_1+(\alpha^2+1)k_2-1]/(\alpha^2+2)}
\cr
&\times&\left[\frac{(z+d)(z_0+d)}{(z-d)(z_0-d)}\right]^{k_2}.
\ee
In this interval, we note that $g_{22}\sim {\cal O}(\rho^2)$, and the rod is in $\partial/\partial x^2$ direction. So we study the period $\triangle x^2$ to find that
\be
\triangle x^2=2\pi \lim_{\rho \to 0}\sqrt{\frac{\rho^2f}{g_{22}}}&=2\pi
\left(\displaystyle{\frac{z_0+d}{z_0-d}}\right)^{k_2/2},\label{conical}
\ee
which shows that the parameters should be adjusted to avoid the conical singularity.\\
\noindent
(iii) $d>z>-d$
\be
g_{00}&\sim&-\left(\frac{d+m}{2d}\right)^{-4/(\alpha^2+2)}
[4(d^2-z^2)]^{(2\alpha^2k_1+\alpha^2k_2-2)/(\alpha^2+2)} \cr
&\times&\rho^{-2(2\alpha^2k_1+\alpha^2k_2-2)/(\alpha^2+2)},\\
g_{11}&\sim&-2(z+z_0)\left(\frac{d+m}{2d}\right)^{2/(\alpha^2+2)}
[4(d^2-z^2)]^{-[\alpha^2k_1+(\alpha^2+1)k_2-1]/(\alpha^2+2)} \cr
&\times&\rho^{2[\alpha^2k_1+(\alpha^2+1)k_2-1]/(\alpha^2+2)},\\
g_{22}&\sim&-\frac{1}{2(z+z_0)}\left(\frac{d+m}{2d}\right)^{2/(\alpha^2+2)}
[4(d^2-z^2)]^{-(\alpha^2k_1-k_2-1)/(\alpha^2+2)} \cr
&\times&\rho^{2(\alpha^2k_1-k_2+\alpha^2+1)/(\alpha^2+2)},\\
f&\sim&-\frac{1}{2(z+z_0)}\left(\frac{d+m}{2d}\right)^{2/(\alpha^2+2)}
[4(d^2-z^2)]^{-[\alpha^2k_1+(\alpha^2+1)k_2-1]/(\alpha^2+2)}\cr
&\times&\left[\frac{4(d^2-z^2)(z+z_0)}{z_0-d}\right]^{k_2}
\left(\frac{d}{d^2-z^2}\right)^{2\nu}
\rho^{2(\alpha^2k_1-k_2-1)/(\alpha^2+2)+2\nu}.
\ee
The behavior in this interval depends on the values of $\alpha$, $k_1$ and $k_2$. \\
\noindent
(iv) $-d>z>-\infty$
\be
g_{00}&\sim&-\left(\frac{z-m}{z-d}\right)^{-4/(\alpha^2+2)}
\left(\frac{z+d}{z-d}\right)^{-(2\alpha^2k_1+\alpha^2k_2-2)/(\alpha^2+2)},\\
g_{11}&\sim&-2(z+z_0)\left(\frac{z-m}{z-d}\right)^{2/(\alpha^2+2)}
\left(\frac{z+d}{z-d}
\right)^{[\alpha^2k_1+(\alpha^2+1)k_2-1]/(\alpha^2+2)},\\
g_{22}&\sim&-\frac{\rho^2}{2(z+z_0)}
\left(\frac{z-m}{z-d}\right)^{2/(\alpha^2+2)}
\left(\frac{z+d}{z-d}\right)^{(\alpha^2k_1-k_2-1)/(\alpha^2+2)},\\
f&\sim&-\frac{1}{2(z+z_0)}\left(\frac{z-m}{z-d}\right)^{2/(\alpha^2+2)}
\left(\frac{z+d}{z-d}\right)^{(\alpha^2k_1-k_2-1)/(\alpha^2+2)}.
\ee
In this interval, we note that $g_{22}\sim {\cal O}(\rho^2)$ as in the interval (ii), and the period $\triangle x^2$ is given by
\be
\triangle x^2=2\pi \lim_{\rho \to 0}\sqrt{\frac{\rho^2f}{g_{22}}}&=2\pi,
\ee
which shows that there is no conical singularity.

%%%%%%%%%%%%%%%%%%%%%%dilaton black ring solution%%%%%%%%%%%%%%%%%%%%%%
We first determine the values of the parameters of $k_1$ and $k_2$ by limiting ourselves to finding physically interesting solutions. We investigate the black hole solutions where the metric components should behave in the interval (iii) as $g_{00}\sim \rho^2,~g_{11}\sim \rho^0,~g_{22}\sim \rho^0,~f\sim \rho^0$. This turns out to require that
\be
\alpha^2(2k_1+k_2+1)&=&0,\\
\alpha^2k_1+(\alpha^2+1)k_2-1&=&0,\\
\alpha^2k_1-k_2+\alpha^2+1&=&0,
\ee
that determine the parameters as
\be
\{\alpha^2=0,~k_2=1\}, \quad {\rm or} \quad \{k_1=-1,~k_2=1\}.
\ee
Noting that  the solution with $\alpha^2=0$ is reduced to the 5D EM solution, we fix $k_1$ and $k_2$ as $k_1=-1,~k_2=1$. Then, the rod in 
$[-d,~d]$ represents the event horizon and is sandwiched by the rods in the same $\partial /\partial x^2$ direction. This shows that the topology of the event horizon is $S^2 \times S^1$ as in the static black ring and static charged black ring cases. Therefore, the solution in 
Eqs.~(\ref{2solitong00})-(\ref{2soliton}) with $k_1=-1$ and $k_2=1$ is a black ring solution with electric and dilatonic charges. The solution can be written as  
\be
g_{00}&=&-K^{-1}h,\label{g00K}\\
g_{11}&=&\mu^*\sqrt{K},\\
g_{22}&=&-\mu\sqrt{K}h^{-1},\label{g22K}\\
f&=&-\frac{(z_0+d)\sqrt{K}\mu\rho^2(\mu_2-\mu_1)^2}
{\sqrt{(z+z_0)^2+\rho^2}(\mu_1-\mu)(\mu_2-\mu)
(\mu_1^2+\rho^2)(\mu_2^2+\rho^2)},
\ee
and dilaton fields and the electric potential are given by
\be
\Phi&=&-\frac{3\alpha}{8}\ln K,\\
\chi&=&\frac{e}{2d}(1-h)K^{-(\alpha^2+2)/4}.
\ee
Note that when $m=d$ and accordingly $e=0$ and $K=1$, the solution is reduced 
to the static black ring solution. 

%%%%%%%%%%%%%%%%prolate spheroidal coordinates%%%%%%%%%%%%%%%%%%%%%%%%%
In order to elucidate the structure of the solution given here, we introduce the prolate spheroidal coordinates $x$ and $y$ by
\be
\rho&=&d\sqrt{(x^2-1)(1-y^2)},\\
z&=&dxy.
\ee
In these coordinates we have the expressions
\be
h&=&\frac{x-1}{x+1},\\
K&=&\left[\frac{dx+m}{d(x+1)}\right]^{4/(\alpha^2+2)},\\
\mu_1&=&d(x-1)(1-y),\\
\mu_2&=&-d(x-1)(1+y),\\
\mu&=&-(dxy+z_0)-\sqrt{(dxy+z_0)^2+d^2(x^2-1)(1-y^2)},\\
\mu^*&=&-(dxy+z_0)+\sqrt{(dxy+z_0)^2+d^2(x^2-1)(1-y^2)}.
\ee
Using these quantities, we can rewrite the metric components $g_{00}$, 
$g_{11}$ and $g_{22}$ in terms of $x$ and $y$, and we also obtain the components $g_{xx}$ and $g_{yy}$ in the metric: 
\be
g_{xx}&=&\frac{-d^2(z_0+d)\sqrt{K}\mu}
{\sqrt{(dxy+z_0)^2+d^2(x^2-1)(1-y^2)}(\mu_1-\mu)(\mu_2-\mu)},\\
g_{yy}&=&\frac{-d^2(z_0+d)\sqrt{K}\mu(x^2-1)}
{\sqrt{(dxy+z_0)^2+d^2(x^2-1)(1-y^2)}(\mu_1-\mu)(\mu_2-\mu)(1-y^2)}. \ \ \
\ee
The point $z=-z_0$ on the $z$-axis in the coordinates $(\rho,z)$ corresponds to $(-z_0/d,1)$ or $(z_0/d,-1)$ in the coordinates $(x,y)$. 
The horizon appearing in the interval $[-d,d]$ on the $z$-axis exists in the region specified by $x=1$ and $-1<y<1$ in $(x,y)$. The region specified by $-1<x<1$ and $-1<y<1$, which can not be expressed by the coordinates $(\rho,z)$, represents the inside of the horizon. We find that a curvature singularity lies at $x=-1$, which can be shown from the curvature invariant behaving as
\be
R^{\alpha\beta\gamma\delta}R_{\alpha\beta\gamma\delta}\sim{\rm constant}\times\frac{z_0-dy}{(x+1)^6}.
\ee
This curvature singularity is covered with the horizon of the dilaton black ring.
%%%%%%%%%%%%%%%%%%%%%%%%%Figure1%%%%%%%%%%%%%%%%%%%%%%%%%%%%%%%%%
\begin{figure}
\centerline{\includegraphics[width=7cm,height=7cm]{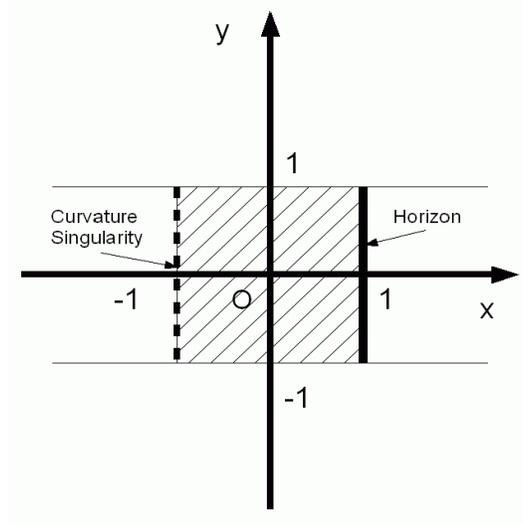}}
\caption{The horizon and the curvature singularity are depicted in the $x$-$y$ plane.  
The hatched region characterized by $-1<x<1$ and $-1<y<1$ is inside of the horizon at $x=1$.}
\label{fig:1}
\end{figure}
%%%%%%%%%%%%%%%%%%%%%%%%%%%%%%%%%%%%%%%%%%%%%%%%%%%%%%%%%%%%%%%%%%%%%%%

%%%%%%%%%%%%%%%%%%%%%%asymptotic behaviors%%%%%%%%%%%%%%%%%%%%%%%%%%%%%
We next study the asymptotic behaviors of the metric components, the dilaton fields and the electric potential. Using 
\be
h &\sim& 1-\frac{2d}{\sqrt{\rho^2+z^2}},\\
K &\sim& 1+\frac{4}{\alpha^2+2}\frac{m-d}{\sqrt{\rho^2+z^2}},
\ee
as $\sqrt{\rho^2+z^2}\to \infty$, we have
\be
&&g_{00}\sim -1,\quad g_{11}\sim -z+\sqrt{\rho^2+z^2},\cr
&&g_{22}\sim z+\sqrt{\rho^2+z^2},\quad
f\sim \frac{1}{2\sqrt{\rho^2+z^2}},
\ee
which shows the metric becomes asymptotically flat, and the fields approach to
\be
\Phi &\sim& -\frac{3\alpha}{2(\alpha^2+2)}\frac{m-d}{\sqrt{\rho^2+z^2}},\\
\chi &\sim& \frac{e}{\sqrt{\rho^2+z^2}}.
\ee

%%%%%%%%%%%%%%%%%removal of conical singularity 1%%%%%%%%%%%%%%%%%%%%%%%
From Eq.~(\ref{conical}) we find that there is a conical singularity in 
$[d, -z_0]$. In order to get rid of the conical singularity, there can be two ways. The first is to set $d=-z_0$, which gets rid of the interval itself. The second is to make $\triangle x^2= 2\pi$ by imposing $d=0$. 
We first compute the $d=-z_0$ case. In this case, the metric components 
$g_{00}, g_{11}$ and $g_{22}$ are given by the same forms as in 
Eqs.~(\ref{g00K})-(\ref{g22K}) with
\be
h=-\frac{\mu_1\mu}{\rho^2},
\ee
and the component $f$ is given by
\be
f=\frac{\sqrt{K}\rho^2(\mu_1-\mu)}
{(\mu_1^2+\rho^2)(\mu+\rho^2)},
\ee
where we have used the limit
\be
\lim_{d\to -z_0}\frac{z_0+d}{\mu_2-\mu}=-\frac{\sqrt{(z+z_0)^2+\rho^2}}{\mu}.
\ee
Introducing the hyper-spherical coordinates $r$ and $\theta$ by
\be
\rho&=&\frac{1}{2}\sqrt{(r^2-2m)^2-4d^2}\sin 2\theta,\\
z&=&\frac{1}{2}(r^2-2m)\cos 2\theta,
\ee
we have
\be
\mu&=&-(r^2-2m-2d)\cos^2 \theta,\\
\mu^*&=&(r^2-2m+2d)\sin^2 \theta,\\
\mu_1&=&(r^2-2m-2d)\sin^2 \theta,\\
h&=&\frac{r^2-2m-2d}{r^2-2m+2d},\\
K&=&\left(\frac{r^2}{r^2-2m+2d}\right)^{4/(\alpha^2+2)}.
\ee
By using these expressions the metric components are reduced to
\be
g_{00}&=&-\frac{r^2-2m-2d}{r^2-2m+2d}\left(\frac{r^2-2m+2d}{r^2}
\right)^{4/(\alpha^2+2)},\\
g_{11}&=&(r^2-2m+2d)\sin^2\theta\left(\frac{r^2}{r^2-2m+2d}
\right)^{2/(\alpha^2+2)},\\
g_{22}&=&(r^2-2m+2d)\cos^2\theta\left(\frac{r^2}{r^2-2m+2d}
\right)^{2/(\alpha^2+2)},\\
f&=&\frac{r^2-2m+2d}{(r^2-2m)^2-4d^2\cos^22\theta}
\left(\frac{r^2}{r^2-2m+2d}\right)^{2/(\alpha^2+2)}.
\ee
Then, noting the relation
\be
d\rho^2+dz^2=\left[(r^2-2m)^2-4d^2\cos^22\theta\right]
\left[\frac{r^2dr^2}{(r^2-2m)^2-4d^2}+d\theta^2\right],
\ee
we obtain
\be
ds^2=&-&\left[1-\frac{2(m+d)}{r^2}\right]
Y^{-(\alpha^2-2)/(\alpha^2+2)}(dx^0)^2 \cr
&+&\left[1-\frac{2(m+d)}{r^2}\right]^{-1}Y^{-2/(\alpha^2+2)}dr^2 \cr
&+&Y^{\alpha^2/(\alpha^2+2)}r^2
[d\theta^2 +\sin^2 \theta(dx^1)^2+\cos^2 \theta(dx^2)^2],
\ee
where
\be
Y=1-\frac{2(m-d)}{r^2}.
\ee
This metric has an event horizon at $r^2=2(m+d)$ and a curvature singularity at $r^2=2(m-d)$. Because this singularity is covered with the horizon, the metric represents 5D dilaton black hole. This type of solution was first given by Gibbons and Maeda\cite{GM}. The difference from their solution comes from the imposition of the coordinate condition $\det g=-\rho^2$ in our solution. We find
\be
g_{00}\cdot g_{rr}=-Y^{-\alpha^2/(\alpha^2+2)},
\ee
which does not equal to $-1$ showing the violation of the equivalence principle. 
%The second possibility of removing the conical singularity is setting%
%$d=0$.% 
The extremal case obtained in the limit $d=0$ has no event horizon but curvature singularity at $r^2=2m$. This should be compared with the 5D EM case where the metric becomes the 5D Majumdar-Papapetrou-type solution. 

%%%%%%%%%%%%%%%%%removal of conical singularity 2%%%%%%%%%%%%%%%%%%%%%%%
The second possibility of removing the conical singularity is setting $d=0$. When $d=0$ and $z_0 \ne 0$, we get the metric components
\be
g_{00}&=&-K_0^{-1},\\
g_{11}&=&\mu^*\sqrt{K_0},\\
g_{22}&=&-\mu\sqrt{K_0},\\
f&=&\frac{\sqrt{K_0}}{2\sqrt{(z+z_0)^2+\rho^2}},
\ee
where
\be
K_0=\left(1+\frac{m}{\sqrt{\rho^2+z^2}}\right)^{4/(\alpha^2+2)}.
\ee
We find that the solution has a ring-like structure at $z=0$, and near the ring the metric components behave as
\be
g_{00} &\sim& \rho^{4/(\alpha^2+2)},\\
g_{22} &\sim& \rho^{2(\alpha^2+1)/(\alpha^2+2)}.
\ee
This shows that there are two eigenvectors belonging to the zero eigenvalue for $g$, which leads to a curvature singularity at $\rho=0$ and $z=0$. 

%%%%%%%%%%%%%%%%%%%%%%%%%%%%%%%%%%%%%%%%%
%
%    section 4
%
%%%%%%%%%%%%%%%%%%%%%%%%%%%%%%%%%%%%%%%%%
\section{Summary and Discussion}

In this paper we obtain an infinite number of static solutions to the 5D EMd equations, which shows that 5D EMd equations are completely integrable despite their apparent complexity. The solutions are specified by their soliton numbers. We examined the 2-soliton solution in detail by studying its rod structure. By requiring the existence of the horizon, we determined the parameters which come in when we assume an ansatz. The rod structure of the 2-soliton solution thus obtained is the same as those of the static black ring case and the static charged black ring case, that is, there is a conical singularity. 

We get rid of the conical singularity by imposing some constraints between the positions of the poles. We thus obtain the 5D dilaton black hole solution. This solution is a comprehensive solution in a sense that the EM solution is obtained by setting $\alpha=0$ and the static black ring solution is obtained by setting $m=d$. The difference between the EMd  black hole solution and EM  black hole solution arises in the extremal case where $d=0$. In the EM case, we have a 5D Majumdar-Papapetrou-type solution. On the other hand, EMd solution has a curvature singularity and so it is not a black hole solution. The same situation can be also found in the 4D case.

It is well known that the Majumdar-Papapetrou(MP) solution is expressed in terms of the multi black hole solution in four dimensions, but, in the 5D case, such a solution does not exist in the EM case, and neither in the present case with the dilaton field. In the 4D case, we can obtain the MP solution by imposing the extremal condition, which enables us to remove the conical singularities aligned along the $z$-axis. The corresponding condition in the present case is $d=0$, by which naked singularities appear in replacement of the vanishing of conical singularities. A possible solution of evading conical singularities might be found in the stationary case such as black saturn$\cite{EF}$ and there may be a dilatonic black saturn solution in the stationay case.  
        
Last of all we list up the subjects which should be studied in future. In addition to the stationary case mentioned above, the higher number of soliton solutions need be investigated to clarify their physical meaning.  Since our construction of the solutions depend on the ansatz (\ref{ansatz1}) and (\ref{ansatz2}), by removing these ansatz and allowing for an arbitrary soliton numbers for $\ln N_2$, $\ln N$ and $\ln h$, new type of physically interesting solutions may be found. In the construction of $n$-soliton solution, we start with a vacuum solution in this paper. When we adopt more complicated solution as the seed solution, we might obtain some other interesting solutions.


\begin{thebibliography}{99}
\bibitem{ER}R.~Emparan and H.~A.~Reall,~Phys.~Rev.~Lett.~{\bf 88}
(2002),~101101.
\bibitem{Pom}A.~A.~Pomeransky,~Phys.~Rev.~D~{\bf 73}
(2006),~044004.
\bibitem{MI}T.~Mishima and H.~Iguchi,~Phys.~Rev.~D~{\bf 73}
(2006),~044030.
\bibitem{TMY}S.~Tomizawa,~Y.~Morisawa and Y.~Yasui,~Phys.~Rev.~D~{\bf 73}
(2006), 064009.
\bibitem{Y1}S.~S.~Yazadjiev,~Phys.~Rev.~D~{\bf 73}
(2006),~104007.
\bibitem{TN}S.~Tomizawa and M.~Nozawa,~Phys.~Rev.~D~{\bf 73}
(2006),~124034.
\bibitem{IM}H.~Iguchi and T.~Mishima,~Phys.~Rev.~D~{\bf 74}
(2006),~024029.
\bibitem{TIM}S.~Tomizawa,~H.~Iguchi and T.~Mishima,~Phys.~Rev.~D~{\bf 74}
(2006),~104004.
\bibitem{Y2}S.~S.~Yazadjiev,~JHEP~{\bf 0607}
(2006),~036.
\bibitem{EF}H.~Elvang and P.~Figueras,~JHEP~{\bf 0705}
(2007),~050.
\bibitem{BS}V.~A.~Belinskii and V.~E.~Sakharov,~Sov.~Phys.~JETP~{\bf 50} 
(1979),~1.
\bibitem{K}T.~Koikawa,~Prog.~Theor.~Phys.~{\bf 114}
(2005),~793. 
\bibitem{AK1}T.~Azuma and T.~Koikawa,~Prog.~Theor.~Phys.~{\bf 116} (2006),~319.
\bibitem{AK2}T.~Azuma and T.~Koikawa,~Prog.~Theor.~Phys.~{\bf 118}
(2007),~35.
\bibitem{AK4D1}T.~Azuma and T.~Koikawa,~Gen.~Rel.~Grav.~{\bf 24}
(1992),~1223.
\bibitem{AK4D2}T.~Azuma and T.~Koikawa,~Prog.~Theor.~Phys.~{\bf 90} (1993),~585.
\bibitem{AK4D3}T.~Azuma and T.~Koikawa,~Prog.~Theor.~Phys.~{\bf 90} (1993),~991.
\bibitem{AK4D4}T.~Azuma and T.~Koikawa,~Prog.~Theor.~Phys.~{\bf 92}  (1994),~1095.
\bibitem{AK4D5}T.~Azuma and T.~Koikawa,~Prog.~Theor.~Phys.~{\bf 93} (1995),~1021.
\bibitem{E}R.~Emparan,~JHEP~{\bf 0403}
(2004),~064.
\bibitem{KL}H.~K.~Kunduri and J.~Lucietti,~Phys.~Lett.~B~{\bf 609} 
(2005),~143.
\bibitem{SR}A.~Sheykhi and N.~Riazi,~Int.~J.~Mod.~Phys.~{\bf 22}
(2007),~4849.
\bibitem{GM}G.~W.~Gibbons and K.~Maeda,~Nucl.~Phys.~B~{\bf 298}
(1988),~741.
\end{thebibliography}
\end{document}